\documentstyle[epsfig,multicol,prb,aps]{revtex}
\begin{document}
\title{Mechanical relaxation in glasses and at the glass transition}
\author {U. Buchenau}

\address {Institut f\"ur Festk\"orperforschung, Forschungszentrum
        J\"ulich, Postfach 1913, 52425 J\"ulich, Germany}
\date{\today}
\maketitle
\begin{abstract}
The Gilroy-Phillips model of relaxational jumps in asymmetric
double well potentials, developed for the Arrhenius-type secondary
relaxations of the glass phase, is extended to a formal
description of the breakdown of the shear modulus at the glass
transition, the $\alpha$-process. The extension requires the
introduction of two separate parts of the barrier distribution
function $f(V)$, with a different temperature behavior of primary
and secondary parts, respectively. The time-temperature scaling of
the $\alpha$-process, together with a sum rule for the whole
barrier distribution function, implies a strong rise of the
integrated secondary relaxation with increasing temperature above
the glass transition. Thus one gets for the first time a
quantitative relation between the fragility of the glass former
and the fast rise of the picosecond process observed in neutron
and Raman scattering.

The formalism is applied to literature data of polystyrene,
vitreous silica and a sodium silicate glass. In the glass phase of
polystyrene, one finds a temperature-independent secondary barrier
distribution function, in agreement with an earlier Raman result
from the literature. Above the glass transition, the secondary
barrier distribution function increases with temperature as
predicted. The findings allow for a new interpretation of the
fragility and the entropy crisis at the glass transition.
\end{abstract}


\begin {multicols} {2}
\narrowtext

\section{Introduction}

Relaxation in glasses, sometimes also called secondary relaxation
to distinguish it from the primary relaxation at the glass
transition, is generally believed \cite{arnold,hunk,heijboer} to
be well described in terms of the Arrhenius-Kramers picture
\cite{kramers}, with a relaxation time $\tau_V$ given by the
Arrhenius relation
\begin{equation} \tau_V=\tau_0{\rm e}^{V/k_BT}, \label{arr}
\end{equation}
where $\tau_0$ is a microscopic time of the order of 10$^{-13}$
seconds, $V$ is the energy of the barrier between two energy
minima of the system, and $T$ is the temperature.

In contrast, the primary relaxation or $\alpha$-process, the onset
of the flow process at the glass transition temperature $T_g$ and
above, seems to follow a much steeper law \cite{ang,edi}
\begin{equation}
\tau_\alpha=\tau_0{\rm e}^{A/(T-T_0)}, \label{vogel}
\end{equation}
where $A$ and $T_0$ are constants with the dimension of a
temperature. This is the well-known empirical Vogel-Fulcher-Tamman
(VFT) or Williams-Landel-Ferry (WLF) equation. $T_0$, the
Vogel-Fulcher temperature, is smaller than $T_g$; the closer it
lies to $T_g$, the more fragile is the glass former.

Since the Arrhenius law has a sound microscopic background
\cite{kramers} and the VFT or WLF equation has not, it seems
reasonable to build a joint quantitative description on the
former, bearing in mind the physical difference of the two
processes. This is the intention of the present paper.

In glasses, one has to reckon with a whole distribution of
relaxational jumps, not only over different potential barrier
heights, but also between energy minima of different energy. Thus
one has to extend the classical Arrhenius-Kramers treatment
\cite{kramers} of a thermally activated relaxation process in a
symmetric double-well potential to deal with a broad distribution
of barrier heights and asymmetries. A distribution in the barrier
heights was first considered by Fr\"ohlich \cite{froh}, but only
for symmetric potentials. An asymmetric multiminimum situation was
solved by Hoffman and Pfeiffer \cite{hoff}. But none of these
early attempts dealt simultaneously with a broad distribution in
both quantities, the barrier height and the asymmetry of the
wells. The necessity of such a double distribution was first
recognized in the tunneling model for the two-level states below 1
K \cite{tunn} in 1971. Ten years later, Gilroy and Phillips
\cite{gilroy} extended the scheme to classical relaxation
processes. Also, they were the first to draw a parallel between
mechanical relaxation data and the quasielastic part of the Raman
scattering. At the end of the same decade, the soft potential
model postulated a relation between the tunneling states and the
low barrier classical relaxation processes in glasses
\cite{parshin}. Nevertheless, up to now only few checks of these
postulates for dynamical mechanical \cite{gilroy,keil,tiel,ramos},
Raman \cite{gilroy,surov,wieder} and neutron \cite{bu88} data have
been reported in the literature.

The present paper begins in Section II with a discussion of the
Gilroy-Phillips model, and a derivation of its connection to
rheology. It turns out that one can define a barrier distribution
function $f(V)$ to describe the mechanical shear relaxation at
different temperatures and frequencies. The integral of this
barrier distribution function over all barrier heights $V$ must
equal 1 to bring the shear modulus down to zero. In order to
include the flow process into the same scheme, one separates
$f(V)$ into two parts, $f_s(V)$ and $f_\alpha(V)$. The first of
these describes the secondary relaxations in the glass phase, the
second describes the $\alpha$-process in the undercooled liquid,
respectively. As will be seen, the sum rule for the total barrier
distribution function supplies a quantitative basis for Angell's
conjecture \cite{angnew} of a relation between the fragility and
the rise of the fast picosecond process above the glass
transition. The low-barrier part of $f_s(V)$ determines the
relaxational part of neutron and Raman scattering. The
corresponding equations are derived.

Section III applies the equations to determine $f(V)$ from
literature data for amorphous polystyrene, vitreous silica and a
sodium silicate glass. From the results, one gets a first
impression as to whether one gets the same secondary barrier
distribution function from different methods, in particular if one
compares the high frequency neutron, Raman and Brillouin
scattering results with the low frequency torsion pendulum or
creep data. Furthermore, one gets a feeling for the amount of
reduction of the shear modulus by the secondary relaxation
processes in the glass phase. As we will see, the findings suggest
a generalized Maxwell criterion for the onset of the glass
transition, namely that the flow begins when the shear response
from the secondary relaxation equals the elastic one. Section IV
compiles and discusses these results and their possible
significance for our view of the glass transition. Summary and
conclusions are given in Section V.

\section{The Gilroy-Phillips model}

\subsection{The asymmetric double-well potential}

Let us denote the shear strain by $\epsilon$, the shear stress by
$\sigma$ and the (infinite frequency) shear modulus by $G$. $G$
will generally depend on the temperature $T$.

\begin{figure}[b]
\begin{center}
\vspace*{-0.5cm}
\hspace*{0.0ex}\epsfig{file=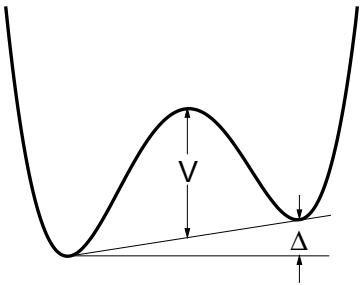,height=8.5cm,angle=-90}

\vspace*{-0.5cm}
\end{center}
\caption{Asymmetric double well potential with barrier height $V$
and asymmetry $\Delta$ as a function of a generalized coordinate.}

\end{figure}

The structural relaxation is taken to be a superposition of
independent Debye relaxation centers in asymmetric double-well
potentials with two minima, as shown in Fig. 1. The energy of the
left minimum is $-\Delta/2$ and the energy of the right minimum is
$+\Delta/2$. The height of the barrier is $V$.

The interaction between the shear strain and the Debye relaxation
center is described by the change of the asymmetry $\Delta$ under
the influence of the strain. The interaction is characterized by
the coupling parameter $\gamma$, leading to an asymmetry
$\Delta+\gamma\epsilon$ of the relaxation in the strained glass.
$\gamma$ must be considered to depend both on $V$ and $\Delta$.

The free energy $F$ of the relaxation center reads
\begin{equation}
F =-k_BT\ln\left[2\cosh\left(\frac{\Delta+\gamma\epsilon}{2k_BT}
\right)\right],
\end{equation}
which has the second derivative with respect to the shear distortion
$\epsilon$
\begin{equation}\label{asymm}
\frac{\partial^2F}{\partial\epsilon^2}
 =-\frac{\gamma^2}{4k_BT\cosh^2(\Delta/2k_BT)}
\end{equation}
The second derivative determines the contribution of that specific
relaxing entity to the difference between the shear moduli at
infinite and zero frequency. The equation shows that the main
influence on the shear modulus is due to relaxation in potentials
with asymmetries smaller than $k_BT$; for larger asymmetries the
influence decreases rapidly because of the square of the
hyperbolic cosine in the denominator.

\subsection{The barrier distribution function $f(V)$}

We want to calculate the frequency dependence of the shear modulus
under the assumption of slowly varying distribution functions in
the parameters $V$ and $\Delta$. In detail, we assume a number
density of relaxing entities $n(V,\Delta)$ and a coupling constant
$\gamma(V,\Delta)$ which are both approximately constant if either
$V$ or $\Delta$ is varied by an amount of the order of the thermal
energy $k_BT$.

Under this assumption, it is safe to neglect as well the influence
of the asymmetry on the relaxation time. We assume the relaxation
time $\tau_V$ to be given by the Arrhenius equation (\ref{arr}).

We then integrate over the asymmetry $\Delta$ to obtain the step
$\delta G$ between the shear moduli at infinite and zero frequency
from all relaxation centers with barrier height between $V$ and
$V+dV$
\begin{equation}
\delta
G=dV\int_{-\infty}^\infty\frac{\gamma^2n(V,\Delta)d\Delta}{4k_BT\cosh^2(\Delta/2k_BT)}
\end{equation}
Since one has only contributions in the near neighborhood of
$\Delta=0$, where $n(V,\Delta)\approx n(V,0)$, and since
\begin{equation}
\int_{-\infty}^\infty\frac{d\Delta}{\cosh^2(\Delta/2k_BT)}=4k_BT,
\end{equation}
one finds
\begin{equation} \delta G=\gamma^2n(V,0)dV.
\end{equation}
This is different from a single relaxation in a symmetric
potential, where the step in the modulus is inversely proportional
to the temperature. The physical reason for this difference is
clear: As the temperature rises, relaxation centers with higher
and higher asymmetry begin to contribute to the step in the
modulus. This is an important difference between relaxation in
crystals and relaxation in disordered matter.

The temperature-independent step in the modulus is determined by
the barrier distribution function $f(V)$, defined by
\begin{equation}
f(V) =\frac{\gamma^2n(V,0)}{G}.
\end{equation}
This parameter combination can be argued to remain independent of
temperature, even if $G$ varies with temperature, considering the
relaxing entity as a small misfit region in an elastic medium
\cite{mura}, a misfit region which is able to change the sign of
the misfit by jumping over the barrier. Here, however, this
argument will not be given in detail.

The frequency dependence of the complex shear modulus at the
frequency $\omega$ and the temperature $T$ reads
\begin{equation}
\frac{G'(\omega,T)}{G}= \frac{G_e}{G}+\int_0^\infty f(V)
\frac{dV}{1+\omega^2\tau_V^2}\label{Gp}
\end{equation}
\begin{equation}
\frac{G"(\omega,T)}{G}= \int_0^\infty f(V) \frac{\omega\tau_VdV}
{1+\omega^2\tau_V^2}\label{Gpp},
\end{equation}
where $\tau_V$ is a function of $V$ by the Arrhenius relation eq.
(\ref{arr}), and $G_e$ is the zero frequency modulus after the
decay of all the relaxations in the system.

These two equations describe the real and the imaginary part of
the frequency-dependent shear modulus at all frequencies and
temperatures. As long as one can reckon with a
temperature-independent number of uncoupled relaxation centers,
the barrier distribution function $f(V)$ remains
temperature-independent.

But even if $f(V)$ depends on temperature, it can still be
formally used to characterize the shear relaxation. As shown in
the next subsection, one can rewrite the conventional rheological
expressions \cite{ferry} in terms of $f(V)$. The advantage of the
choice of $f(V)$ lies in the possibility to distinguish the
trivial Arrhenius temperature dependence from other, nontrivial
temperature changes. These nontrivial temperature changes will
then reflect in a temperature dependence of $f(V)$.

\subsection{Rheological equations in terms of $f(V)$}

Comparing the two expressions, equs. (\ref{Gp}) and (\ref{Gpp}),
to those in the textbooks, for instance the one on polymers by
Ferry \cite{ferry} (chapter 3, equs. (23) and (24)), one finds the
relation between the rheological relaxation function $H(\tau)$ and
the barrier distribution function $f(V)$
\begin{equation}\label{rheo}
H(\tau_0{\rm e}^{V/k_BT})=H(\tau_V)=Gk_BTf(V).
\end{equation}

With this equation, one can rewrite all the exact and approximate
rheological relations \cite{ferry} in terms of $f(V)$. To do this,
one first has to define a convenient equivalent function $l(V)$ to
the conventional rheological function $L(\tau)$, which is needed
whenever one wants to calculate a compliance
\begin{equation}
L(\tau_0{\rm e}^{V/k_BT})=L(\tau_V)=\frac{k_BTl(V)}{G}.
\end{equation}

In the following, the most important equations of chapter 3 of
Ferry's book \cite{ferry} are translated into the Gilroy-Phillips
notation. Ferry's equation (19) for the time-dependent modulus
$G(t)$ reads \begin{equation}\label{gt}
G(t)=G_e+\int_{-\infty}^\infty H(\tau) {\rm e}^{-t/\tau}d\ln\tau
\end{equation}
and translates into
\begin{equation} G(t)=G_e+G\int_0^\infty f(V)
{\rm e}^{-t/\tau_V}dV.
\end{equation}
For a viscoelastic liquid, the zero frequency modulus $G_e=0$, so
one must have
\begin{equation}\label{sum} \int_0^\infty f(V)dV=1.
\end{equation}
This is the sum rule for the barrier distribution function $f(V)$.
It has important consequences for the connection between primary
and secondary relaxation, as discussed in the next subsection.

Next, there is Ferry's eq. (20) for the compliance
\begin{equation} J(t)=\frac{1}{G}+\int_{-\infty}^\infty L(\tau)
(1-{\rm e}^{-t/\tau})d\ln\tau+\frac{t}{\eta_0},
\end{equation}
where $\eta_0$ is the viscosity, which translates into
\begin{equation}\label{jt} J(t)=\frac{1}{G}\left[1+\int_0^\infty l(V)(1-{\rm
e}^{-t/\tau_V})dV\right]+\frac{t}{\eta_0}.
\end{equation}

The viscosity $\eta_0$ can be calculated from Ferry's eq. (28)
\begin{equation} \eta_0=\int_{-\infty}^\infty \tau
H(\tau)d\ln\tau,
\end{equation}
which translates into
\begin{equation}\label{eta} \eta_0=G\int_0^\infty \tau_Vf(V)dV.
\end{equation}

Ferry's equs. (21) and (22), the transformation from $H(\tau)$ to
$L(\tau)$ and back, read
\begin{equation}
L=\frac{H}{\left[G_e-\int_{-\infty}^\infty\frac{H(u)}{\tau/u-1}d\ln
u\right]^2+\pi^2H^2}
\end{equation}
and
\begin{equation}
H=\frac{L}{\left[\frac{1}{G}+\int_{-\infty}^\infty\frac{L(u)}{1-u/\tau}d\ln
u-\frac{\tau}{\eta_0}\right]^2+\pi^2L^2}.
\end{equation}

They translate into
\begin{equation}\label{f2l}
\nonumber l(V)=\\ \frac{f(V)}{\left[\frac{G_e}{G}-I_f(V)\right]^2
+(\pi k_BTf(V))^2}
\end{equation}
with
\begin{equation}
I_f(V)=\int_0^\infty\frac{f(E)dE}{\exp((V-E)/k_BT)-1}
\end{equation}
 and
\begin{equation}\label{l2f}
\nonumber f(V)=\\ \frac{l(V)}{\left[1-\frac{\tau_VG}{\eta_0}+
I_l(V) \right]^2+(\pi k_BTl(V))^2}
\end{equation}
with
\begin{equation}
I_l(V)=\int_0^\infty\frac{l(E)dE}{1-\exp((E-V)/k_BT)}.
\end{equation}

With these exact equations, one can calculate the mechanical
response for any type of shear experiment for a given barrier
distribution function $f(V)$. The reverse, the determination of
$f(V)$ from experimental data, is more difficult, because the
exact equations are integral equations. Nevertheless, one can
start to determine a first approximation to $f(V)$ from
measurements of $G'$ and $G"$ using the crude approximations
\begin{equation}\label{gpa}
\frac{G'}{G}= 1-\int_0^{k_BT\ln(1/\omega\tau_0)}f(V)dV \label{G'}
\end{equation}
and
\begin{equation}
f( k_BT\ln(1/\omega\tau_0))=\frac{2}{\pi}\frac{G"}{G k_BT}.
\label{fvexp}
\end{equation}

\subsection{Primary and secondary relaxation}

It is quite clear that one needs to distinguish secondary and
primary processes, because their physical mechanism is different.
Thus one has to distinguish between $f_s(V)$, the secondary
barrier distribution function of the secondary Arrhenius
relaxation, and $f_\alpha(V)$, the primary barrier distribution
function for the primary $\alpha$-process or flow process (see
Fig. 2).

As is well established \cite{ferry}, $H_\alpha(\tau/\tau_\alpha)$
is independent of the temperature. This is the time-temperature
scaling of the $\alpha$-process, sometimes also denoted as
thermorheological simplicity. Since the factor $Gk_BT$ in eq.
(\ref{rheo}) between $H(\tau)$ and $f(V)$ varies only weakly with
temperature, this implies that the primary barrier distribution
function $f_\alpha(V)$ is an essentially temperature-independent
function of $V-V_\alpha(T)$, where $V_\alpha(T)$ denotes the
maximum of this strongly peaked function. From the Vogel-Fulcher
law eq. (\ref{vogel}), one expects the temperature dependence
\begin{equation}\label{valph}
V_\alpha(T)=V_\alpha(T_g)\frac{T(T_g-T_0)}{T_g(T-T_0)},
\end{equation}
showing the divergence of the fictive Arrhenius barrier of the
flow process towards the Vogel-Fulcher temperature $T_0$.

\begin{figure}[b]
\begin{center}
\vspace*{-0.5cm}
\hspace*{0.0ex}\epsfig{file=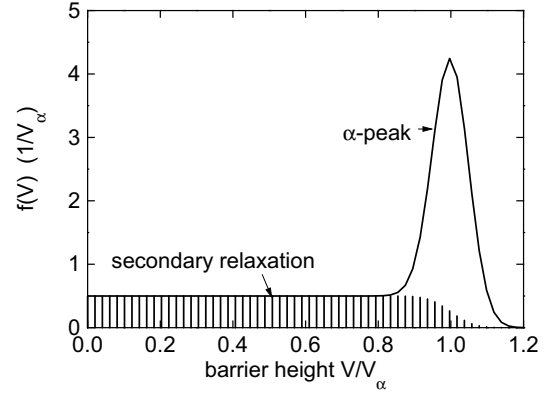,height=8.5cm,angle=-90}

\vspace*{0.0cm}
\end{center}
\caption{Secondary barrier distribution function $f_s(V)$ (the
shaded area), together with the cutoff by the $\alpha$-relaxation
peak (schematic).}

\end{figure}

If time-temperature scaling holds, the weight $w_\alpha$ of the
$\alpha$-process, given by
\begin{equation}
w_\alpha=\int_0^\infty f_\alpha(V)dV
\end{equation}
should be temperature-independent. In the comparison to
experiment, we will see that $w_\alpha$ tends to be close to 1/2.

What does this imply for the secondary relaxations? The
$\alpha$-process is also an upper cutoff for the secondary
relaxation; at the end of the process, the long time shear modulus
is zero. If a secondary relaxation barrier is too high, the
relaxing entity will flow away before it has a chance to jump.
Therefore there is a steep cutoff for the secondary barrier
distribution function $f_s(V)$ at $V_\alpha(T)$. According to the
sum rule eq. (\ref{sum})
\begin{equation}\label{sumfin} \int_0^{V_\alpha(T)}
f_s(V)dV =1- w_\alpha={\rm const}\approx\frac{1}{2}.
\end{equation}

With this sum rule, the decrease of $V_\alpha(T)$ with increasing
temperature implies that the secondary barrier distribution
function  $f_s(V)$ {\it must increase} with increasing
temperature. This increase will be stronger for more fragile glass
formers. Such a connection between the fragility and the rise of
the fast process above $T_g$ has been indeed postulated
empirically \cite{angnew}; here we will be able to quantify this
connection.

The increase of $f_s(V)$ above the glass temperature can be
characterized to first order by
\begin{equation}\label{fsrise}
f_s(V)=f_s(V,T_g)\left[1+\alpha_s(V)\frac{T-T_g}{T_g}\right].
\end{equation}

One can define an average temperature coefficient
$\overline{\alpha_s}$ by
\begin{equation}\label{as}
\overline{\alpha_s}=\frac{1}{1-w_\alpha}\int_0^{V_\alpha(T_g)}\alpha_s(V)f_s(V)dV.
\end{equation}

Differentiating the sum rule eq. (\ref{sumfin}) for $f_s(V)$ with
respect to the temperature, calculating the derivative of
$V_\alpha(T)$ with respect to temperature from eq. (\ref{valph})
and using eq. (\ref{as}), one finds the relation between the
fragility and the average rise of the secondary relaxation
\begin{equation}\label{fin}
\frac{T_0}{T_g-T_0}=\frac{\overline{\alpha_s}(1-w_\alpha)}{V_\alpha(T_g)
f_s(V_\alpha(T_g))}.
\end{equation}

This relation will become clearer in the next subsection, where
the simplest possible case of a constant secondary barrier
distribution function is discussed.

\subsection{The generic case $f_s(V)=$ const}

The deep implications of the Gilroy-Phillips formulation of the
$\alpha$-process are more clearly seen in the simplest possible
case, shown in Fig. 2. Let us assume $f_s(V)={\rm const}$ and
$\alpha_s(V)={\rm const}$. If the rise of $f_s(V)$ with
temperature is strictly linear, equs. (\ref{sumfin}) and
(\ref{fsrise}) imply
\begin{equation}
V_\alpha(T)=\frac{V_\alpha(T_g)T_g}{T_g+\alpha_s(T-T_g)},
\end{equation}
which is not exactly equal, but very close to the empirical
Vogel-Fulcher-Tamman or Williams-Landel-Ferry equation
(\ref{vogel}), with the Vogel-Fulcher temperature $T_0$ given by
\begin{equation}\label{tvtk}
T_0=T_g\left(1-\frac{1}{\alpha_s}\right).
\end{equation}

It is obvious how this comes about: at $T_0$, the density of
secondary processes extrapolates to zero. Thus one has to proceed
to infinitely high barriers to satisfy the sum rule
(\ref{sumfin}). This gives a completely new view on the puzzling
fragility of glass formers: the abnormal temperature dependence is
a consequence of the time-temperature scaling of the
$\alpha$-process, and of a strictly linear rise of the number of
secondary relaxing units with temperature.

The decrease of the secondary barrier distribution function
$f_s(V)$ implies a decrease of the number of minima of the glass
former. Thus one gets an equality \cite{angrao} between the
Vogel-Fulcher and the Kauzmann temperature $T_K$, the latter being
defined as the temperature where the excess entropy of the glass
former over the corresponding crystalline system extrapolates to
zero. If there are no minima between which the glass former can
jump, there is no excess entropy, the old Adam-Gibbs idea
\cite{adam}.

Note that eq. (\ref{tvtk}) for the Vogel-Fulcher or Kauzmann
temperature holds not only in the generic case of a constant
secondary barrier distribution function, but for any $f_s(V)$, as
long as one can reckon with the same strictly linear temperature
rise of the function for all $V$.

To complete the discussion of the glass transition peculiarities
\cite{edi} in the Gilroy-Phillips picture, let us look at the
stretching of the $\alpha$-process, empirically described by the
Kohlrausch equation
\begin{equation}\label{kohlrausch}
G(t)=G_\alpha\exp\left(-(\tau/\tau_\alpha)^\beta\right),
\end{equation}
where the Kohlrausch exponent $\beta$ lies \cite{ang} between 0.3
and 0.7, and $G_\alpha$ is a free parameter. The smaller $\beta$
is, the more stretched is the $\alpha$-relaxation, and the
stronger it deviates from a simple exponential decay.

In order to calculate $\beta$, the definition of the primary
barrier distribution function $f_\alpha(V)$ must be more specific.
Let us assume a gaussian centered at $V_\alpha$ with weight
$w_\alpha=1/2$ and with a full width at half maximum (FWHM)
denoted by $\Delta_\alpha$. Calculating $G(t)$ from eq.
(\ref{gt}), one then finds fairly Kohlrausch-like curves for the
simple generic case of Fig. 2, at least in the time region of the
$\alpha$-process, with $G_\alpha\approx 0.55..0.75\ G$ and
$\tau_\alpha$ about 80 \% of the Arrhenius value for $V_\alpha$.
There is a deviation of the Kohlrausch fits from the calculated
curves, but it is so small that it would be hard to see in an
experiment. It turns out that the ratio $\Delta_\alpha/V_\alpha$
determines the Kohlrausch exponent $\beta$; if it is 0.05, then
$\beta\approx 0.7$; for the ratio 0.1, $\beta\approx 0.5$ and for
the ratio 0.2, $\beta\approx 0.3$. So the broader $f_\alpha(V)$,
the more stretched the relaxation, not unexpected.

Note this is merely a change of description. The Gilroy-Phillips
formulation does not really explain the puzzling features of the
glass transition, the fragility, the entropy crisis and the
stretching. But it supplies a description which allows to look for
a new kind of explanation. We will return to this point in the
discussion.

In order to measure $\alpha_s(V)$ in the picosecond range by
scattering methods, one still needs the equations for the neutron
and Raman scattering functions in terms of the barrier
distribution function. These will be derived in the next
subsection, the last part of the description of the
Gilroy-Phillips model.

\subsection{Neutron and Raman scattering}

One can carry out the same integrations over asymmetries and
barrier heights as in the shear relaxation for the neutron
scattering cross section. Let us begin with a single asymmetric
double well, let us assume that atom $j$ has a coherent scattering
length $b_j$ and an incoherent scattering cross section
$\sigma_j$, and that it jumps from the position $-\vec{d}_j/2$ to
$\vec{d}_j/2$, with the origin of the coordinate system in the
middle between the two.

For the incoherent inelastic scattering in the one-phonon
approximation \cite{marlov}, it suffices to calculate the mean
square displacements. These can be obtained from the Boltzmann
occupation factors of the two minima of the potential. The average
position vector $\vec{r}_j$ of atom $j$ is given by
\begin{equation}
<\vec{r}_j> = -\frac{\vec{d}_j}{2}\tanh{\frac{\Delta}{2k_BT}}
\end{equation}
and its average square is
\begin{equation}
<\vec{r}_j^{\ 2}> = \frac{d_j^2}{4},
\end{equation}
so the mean square displacement contribution of the relaxation to
atom $j$ reads
\begin{equation}
<u_j^2>=<\vec{r}_j^{\ 2}>-<\vec{r}_j>^2
=\frac{d_j^2}{4\cosh^2{\Delta/2k_BT}}.
\end{equation}
In this expression, we recognize the same inverse $\cosh^2{}$ as
in eq. (\ref{asymm}), which can be again integrated over the
asymmetries, if the jump eigenvector stays essentially the same
for the different local asymmetries.

The scattering contribution is a Lorentzian with a half width at
half maximum in $\omega$ given by the inverse relaxation time,
determined by the barrier height according to the Arrhenius
equation (\ref{arr}). The weight of the contribution in
$S(Q,\omega)$ is determined by the number of relaxations and the
weighted sum of the jump vectors
\begin{equation}
d^2=\frac{\sum_j \sigma_j d_j^2}{\overline{\sigma}}.
\end{equation}
The sum is over all atoms in the sample, and $\overline{\sigma}$
is their average incoherent cross section.

Integrating over the barrier heights as well, one gets the
equation for the incoherent scattering
\begin{equation}\label{sqom} S_{inc}(Q,\omega)=v_a
n\left(k_BT\ln{\frac{1}{\omega\tau_0}},0\right)\frac{k_B^2T^2Q^2d^2}{3\omega},
\end{equation}
where $v_a$ is the atomic volume, and the prefactor $1/3$ stems
from the directional average. This equation is again an
approximation, which holds if $f(V)$ does not vary strongly with
$V$. The coherent scattering is obtained replacing $d^2$ by
$d_{coh}^2(Q)$ with
\begin{equation}\label{sqcoh}
d_{coh}^2(Q)=\frac{3}{\overline{b}^2Q^2}
<\mid\sum_jb_j{\rm
e}^{i\vec{Q}\vec{R}_j}\vec{Q}\vec{d}_j\mid^2>,
\end{equation}
where $\vec{R}_j$ is the equilibrium position of atom $j$,
$\overline{b}$ is the average scattering length and the brackets
indicate the orientational averaging over the structure factor of
the relaxation. Even after this averaging, the structure factor
need not show a simple $Q^2$ behaviour like the incoherent one,
but does still contain information on the jump vectors
\cite{bu88}.

The scattering measurements do not give $f(V)$, but rather the
product $n(V,0)d^2$. If one wants the proportionality factor
between those two quantities, one needs additional information
about the relaxing entities. However, there is an elegant and
general way to obtain this proportionality factor for very low
barriers from the soft potential model \cite{parshin}, which
describes the tunneling states and the low-barrier classical
relaxation as similar modes with a double-well potential
distribution. It is not very difficult to derive an equation for
the barrier distribution function $f(V)$ as defined here in terms
of the definitions in this paper. One finds
\begin{equation}\label{spfv}
f_{sp}(V)=\frac{2C}{W^{3/4}V^{1/4}},
\end{equation}
where $C$ (in principle $C_l$ for longitudinal waves and $C_t$ for
transverse waves) is a dimensionless constant of the order of
$10^{-4}$, which can be taken from acoustic attenuation
measurements below 4 K, and $W$ is the crossover energy between
tunneling and vibrational modes, which can be measured from the
crossover regions of the specific heat or the thermal conductivity
at low temperatures \cite{ramos}. The soft potential model has a
fourth parameter $P_s$ for the density of these modes. With this
parameter, the relation between $n(V,0)d^2$ and the secondary
barrier distribution function $f_s(V)$ reads
\begin{equation}\label{spfac}
v_an(V,0)d^2\frac{2W^2\rho C_t}{\hbar^2P_s}=f_s(V),
\end{equation}
where $\rho$ is the mass density. For glasses consisting of more
than one sort of atoms, the relation might fail if strongly and
weakly scattering atoms have different jump widths. But with this
relation, one can determine the secondary barrier distribution
function $f_s(V)$ from neutron scattering measurements without
adaptable parameter, using soft potential parameters from the
literature \cite{ramos}.

The Raman scattering from relaxations in glasses is not so easily
calculated. However, experience \cite{wuttke,sokolov} shows that
neutron and Raman scattering give the same spectra as long as one
stays at frequencies well below the boson peak. Thus one can use
the Raman spectra as one uses incoherent neutron scattering data,
with the disadvantage of an additional general adaptable parameter
for the overall intensity. The advantage of the Raman technique is
a much higher intensity and a much better resolution, allowing one
to assess much lower frequencies.

\section{Comparison to experiment}

\subsection{Values from different techniques}

Let us first consider which barriers one samples with a given
technique. Creep measurements cover the time range from 0.1
seconds to several weeks. This is a measurement in the time
domain, applied mostly to measurements of the $\alpha$-process at
the glass transition, where one should not use the Arrhenius
relation eq. (\ref{arr}). Nevertheless, one can formally calculate
barrier heights of the order of 27 to 41 $k_BT_g$, around 1 eV for
polystyrene ($T_g=373\ K$) and around 4 eV for vitreous silica
($T_g=1473\ K$).

The torsion pendulum method with frequencies around 1 Hz sees
relaxations around 0.1 seconds. In terms of the Arrhenius relation
with $\tau_0=10^{-13}\ s$, this implies a barrier height of 50 meV
at 20 K. In order to get to 1 eV, one has to rise the temperature
by a factor of twenty, thus getting close to the glass temperature
of polystyrene.

Proceeding to higher frequencies, on has the vibrating reed
measurements around 10 kHz, ultrasonic data in the MHz range,
light scattering Brillouin data around 10 GHz and, finally, Raman
and neutron data between a few GHz and a few hundred GHz. For the
latter two, the lower limit holds only for the Raman technique; if
one looks for the weak quasielastic scattering from secondary
relaxations, the neutron technique in practice has a lower limit
of 100 GHz. The upper limit of about 300 GHz is given by the
crossover from relaxational to vibrational scattering
\cite{surov,koizumi}. Thus one sees only the uppermost frequency
band of the relaxational scattering with neutrons. Nevertheless,
neutrons play an important role, because they serve to validate
the Raman scattering data.

For the fast relaxation at 200 GHz, the relaxation time is of the
order of a picosecond, only ten times longer than the microscopic
time scale of the vibrational motion. The Arrhenius relation
translates this again into a barrier of 50 meV at 300 K. Thus
neutron and Raman measurements at room temperature sample the same
relaxations that one sees in a torsion pendulum measurement around
20 K. If one is still in the glass phase at room temperature, one
can thus check the temperature independence of the barrier
distribution function $f(V)$ (in that case, $f(V)=f_s(V)$) by a
comparison between a torsion pendulum and a neutron or Raman
experiment. Naturally, the same can be done by a comparison of a
torsion pendulum and a Brillouin scattering experiment.

If one wants to determine the barrier distribution function from
the mechanical damping for different frequencies, one often has to
compare measurements of different elastic constants. A torsion
pendulum measurement provides immediately the real and imaginary
part of the shear modulus $G$. The vibrating reed technique
measures Youngs modulus $Y$. In terms of the bulk modulus $B$, the
inverse of the compressibility, Youngs modulus reads
\begin{equation}
Y=\frac{9BG}{3B+G}.
\end{equation}
Since that quotient is close to $3G$, it is a reasonable
approximation to identify the $\tan\delta=G'/G"$ results of the
vibrating reed technique (in that technique often denoted as
$Q^{-1}=\tan\delta$) with that of the shear modulus. The real part
of the shear modulus can then be estimated to be $Y'/3$, or
calculated accurately from the above equation if the bulk modulus
is known.

A less safe connection is to longitudinal sound measurements,
which determine the real and imaginary part of the elastic
constant $c_{11}$, sometimes also denoted as the modulus $M$. In
terms of $B$ and $G$, $c_{11}$ reads
\begin{equation}
c_{11}=B+\frac{4}{3}G.
\end{equation}
Both contributions are of comparable size. Therefore, if one wants
to determine $\tan\delta$ for the shear modulus from longitudinal
sound attenuation data, one has to know the relative size of the
two damping contributions from compression and from shear. Here it
should be noted that the mechanical damping at very low
temperatures is quite similar for transverse and longitudinal
phonons in many glasses \cite{berr}. If one believes the soft
potential hypothesis for a common origin of tunneling states and
low barrier relaxation in glasses, the same should hold in the
temperature range up to 50 K. Then one can take $\tan\delta$ from
the inverse mean free path $l^{-1}$ of the longitudinal sound wave
according to
\begin{equation}
\tan\delta=\frac{vl^{-1}}{2\omega},
\end{equation}
where $v$ is the sound velocity.

It is better to use data from transverse sound waves; there, the
damping relates directly to the shear. However, in particular for
light scattering Brillouin experiments, it is much easier to
determine the damping for the longitudinal waves, which provide a
much clearer signal. That damping is often given in terms of the
half width at half maximum $\Gamma$ of the Brillouin line. Then
\begin{equation}
\tan\delta=\frac{\Gamma}{2\omega},
\end{equation}
where $\omega$ is the frequency of the Brillouin line.

Finally, one problem of the determination of $f(V)$ is to know $G$
at the temperature of the measurement; usually one has only $G'$
and $G"$. A way out of this problem is to start at low
temperatures, where $G\approx G'$, determine $f(V)$ via eq.
(\ref{fvexp}), and then pursue $G'/G$ to higher temperatures via
eq. (\ref{gpa}). This way was followed throughout in this paper,
at least as far as the determination of the secondary relaxation
was concerned (the primary relaxation is too sharp to use the
crude approximation of eq. (\ref{fvexp}); there, one has to use
the exact equations).

One can try to check the $G(T)$ values by Brillouin light
scattering measurements of the transverse sound waves, usually at
frequencies around 10 GHz. Even there, one has to reckon with some
influence from the low-barrier part of the relaxations; this is,
however, only a small correction which can be easily done if one
has $f(V)$, using eq. (\ref{G'}).

\subsection{Polystyrene: secondary relaxation}

We begin the comparison to experiment with a heavily studied glass
former, atactic polystyrene, one of the most fragile substances
\cite{ang}, an amorphous polymer where one can rely on a large
number of experimental data, both at low temperatures and at the
glass transition.

The low temperature data were evaluated for temperatures above 10
K; at that limiting temperature, one can begin to reckon with the
validity of the Kramers picture \cite{kramers}. Fig. 3 shows a
compilation of many mechanical low temperature data: torsion
pendulum data at 1 Hz \cite{schwarzl}, at 6 Hz \cite{sinnott},
vibrating reed data at 3, 34 and 87 kHz \cite{nittke,yano,topp}
and Brillouin damping of longitudinal sound waves at 10 GHz
\cite{sonostru}. The data cannot be said to coincide perfectly in
this Gilroy-Phillips evaluation; nevertheless, the agreement is
fair enough to support the concept of a constant number density of
uncoupled relaxation centers. There is no systematic variation
with frequency; one rather has the impression that the differences
stem from the different sample preparation of these seven
measurements. The line in Fig. 3 represents the fit of $f_s(V)$ to
these mechanical data. It is seen that $f_s(V)$ rises towards low
barriers, as one would expect from the soft potential equation
(\ref{spfv}). But the fit line falls below the soft potential
expectation already at rather low barriers, similar to
observations in other glasses \cite{ramos}.

\begin{figure}[b]
\begin{center}
\vspace*{-0.5cm}
\hspace*{0.0ex}\epsfig{file=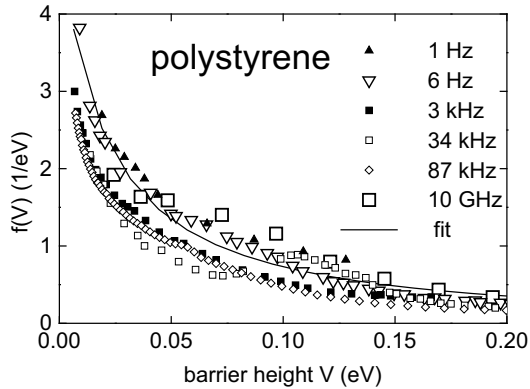,height=8.5cm,angle=-90}

\vspace*{0.0cm}
\end{center}
\caption{Secondary barrier distribution function $f_s(V)$
calculated from literature data of the mechanical damping of
amorphous polystyrene in the glass phase at different frequencies.
References see text. The line is a fit; the same fit is also shown
in Figs. 4, 5 and 6.}

\end{figure}

Fig. 4 compares the same fit to the evaluation of Raman
\cite{surov} and neutron \cite{koizumi} data in terms of eq.
(\ref{sqom}). In the Raman case, data were adapted by an
appropriate multiplication factor. The neutron data were evaluated
using eq. (\ref{spfac}) and the soft potential fit parameters of
polystyrene \cite{ramos}, without any free parameter. The fit of
the neutron data requires a good description of the vibrational
part of the scattering; details will be given in a forthcoming
publication combining time-of-flight and backscattering data of
polystyrene \cite{koizumi}.

The good agreement between data points and the mechanical data fit
line in Fig. 4 corroborates the earlier conclusion \cite{surov} of
a temperature-independent $f_s(V)$ in the glass phase of
polystyrene. Note that the earlier conclusion does not stem from a
comparison of Raman and mechanical data, but rather from a
comparison of Raman data at three different temperatures, namely
100, 200 and 300 K. The two different ways to check the
temperature behaviour of $f_s(V)$ in the glass phase provide the
same result.

\begin{figure}[b]
\begin{center}
\vspace*{-0.5cm}
\hspace*{0.0ex}\epsfig{file=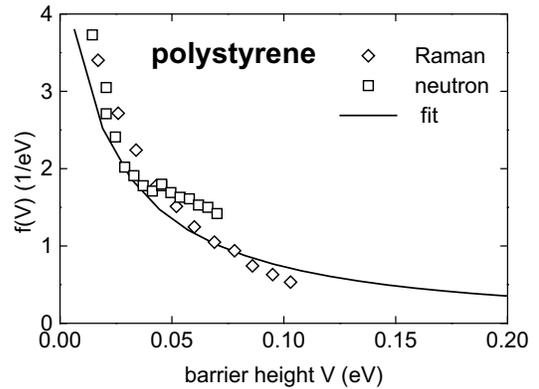,height=8.5cm,angle=-90}

\vspace*{0.0cm}
\end{center}
\caption{Secondary barrier distribution function $f_s(V)$
calculated from literature data of neutron and Raman scattering
from amorphous polystyrene in the glass phase. References see
text.}

\end{figure}

\begin{figure}[b]
\begin{center}
\vspace*{-0.5cm}
\hspace*{0.0ex}\epsfig{file=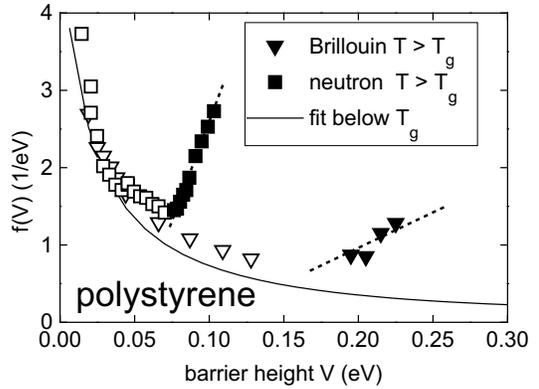,height=8.5cm,angle=-90}

\vspace*{0.0cm}
\end{center}
\caption{Secondary barrier distribution function $f_s(V)$
calculated from literature data of neutron and Brillouin data of
amorphous polystyrene, both in the glass phase and above the glass
temperature $T_g$. References see text.}

\end{figure}

This temperature-independence, however, does no longer hold in the
undercooled liquid phase, above the glass temperature of 372 K of
polystyrene. Fig. 5 shows neutron \cite{koizumi} and longitudinal
sound wave damping data from the Brillouin technique
\cite{patterson} above $T_g$. In order to relate to the preceding
figures, $f(V)$ is again plotted against the barrier height $V$.
This implies that one sees the onset of the increase of $f_s(V)$
with increasing temperature at different values of $V$ in the two
techniques, at 0.074 eV for the neutrons and at 0.163 eV for the
Brillouin data. Note that in both cases the frequency is too high
to see the $\alpha$-process at the temperatures of the
measurement.

We conclude that $f_s(V)$ does indeed increase above $T_g$, as
postulated above on the basis of the sum rule for $f(V)$, eq.
(\ref{sum}), and on the basis of the temperature dependence of the
$\alpha$-process. The rise of $f_s(V)$ above $T_g$ can be
characterized by the linear relation eq. (\ref{fsrise}) with
$\alpha_s=7.7\pm 1$ for the neutron data and $\alpha_s=10\pm 3$
for the Brillouin data (in the latter case, the large error is due
to the small number of points and the insecurity of the value at
$T_g$), within experimental error the same temperature coefficient
for both sets of data.

Fig. 6 shows $f_s(V)$ for polystyrene over the whole barrier
range, together with the results of a torsion pendulum measurement
at 1 Hz up to $T_g$ \cite{schwarzl}, an ultrasonic measurement
just below $T_g$ \cite{piche} and the neutron result up to $T_g$
\cite{koizumi}. The shaded area represents $f_s(V)$; the
$\alpha$-peak above 1 eV will be discussed in the next subsection.

\begin{figure}[b]
\begin{center}
\vspace*{-0.5cm}
\hspace*{0.0ex}\epsfig{file=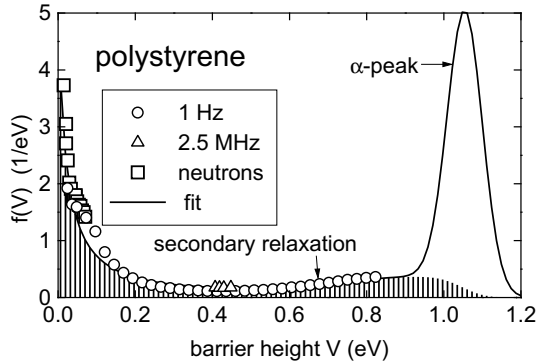,height=8.5cm,angle=-90}

\vspace*{0.0cm}
\end{center}
\caption{Secondary barrier distribution function $f_s(V)$
calculated from literature mechanical damping and neutron data of
amorphous polystyrene up to the glass transition. The peak at the
end shows the gaussian $f_\alpha(V)$ describing the
$\alpha$-process at the glass temperature $T_g$. References see
text.}

\end{figure}

\subsection{Polystyrene: glass transition}

The $\alpha$-process is characterized by the position
$V_\alpha(T)$, the full width at half maximum $\Delta_\alpha$ of
$f_\alpha(V)$ and the weight $w_\alpha$. The latter two should be
temperature-independent. Let us define
$V_\alpha(T_g)=15k_BT_g\ln(10)$ in order to have the corresponding
relaxation time at 100 s, and let us choose a gaussian for the
primary barrier distribution function $f_\alpha(V)$.

If one knows $f_s(V)$ from measurements in the glass phase,
$w_\alpha$ can be calculated from the sum rule eq. (\ref{sumfin}).
In the case of polystyrene, one finds $w_\alpha=0.54$ for the fit
function in Fig. 6.

The width $\Delta_\alpha=0.107$ eV, the glass temperature
$T_g=372\ K$ and the coefficient $\alpha_s=6.9\pm 0.1$ was fitted
to creep data \cite{plazek} of polystyrene with a molecular weight
of 600000 g/mol, data which are also treated in Ferry's book
\cite{ferry} and which are shown in Fig. 7. The fit requires as an
additional parameter the infinite frequency modulus $G=1.69$ GPa
at 373.8 K. Together with the low temperature data and the known
density variation of polystyrene with temperature
\cite{schwarzl1}, one calculates a Gr\"uneisen $\Gamma_g=3.8$ for
$G$ in polystyrene. With this value, one can also calculate $G$
for higher temperatures. The creep function $J(t)$ is obtained by
first calculating $l(V)$ via eq. (\ref{f2l}), the viscosity from
eq. (\ref{eta}) and finally $J(t)$ from eq. (\ref{jt}).

\begin{figure}[b]
\begin{center}
\vspace*{-0.5cm}
\hspace*{0.0ex}\epsfig{file=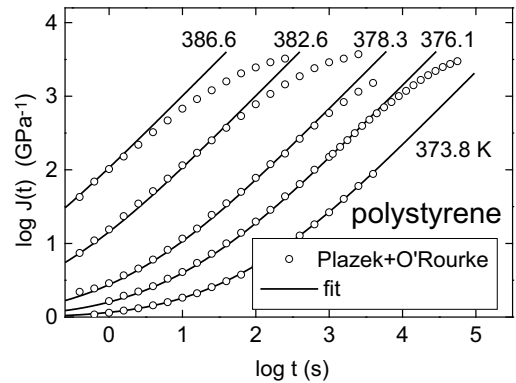,height=8.5cm,angle=-90}

\vspace*{0.0cm}
\end{center}
\caption{Creep data of the $\alpha$-process in polystyrene at the
glass transition (reference see text), together with a fit in
terms of $f(V)$.}

\end{figure}

As seen in Fig. 7, one can describe the temperature shift of the
$\alpha$-process with an appropriate rise of $f_s(V)$ above $T_g$.
The sum rule eq. (\ref{sumfin}) forces a temperature shift of
$V_\alpha(T)$, which in turn provides the experimentally observed
shift factors towards higher temperatures.

The scheme works quite well up to three decades in compliance; for
still higher compliances, one gets into the plateau regime from
the chain entanglement \cite{ferry}, which is beyond the present
considerations.

\subsection{Vitreous silica and sodium silicate}

Vitreous silica is the case for which the Gilroy-Phillips idea
\cite{gilroy} was originally developed. The compatibility between
mechanical relaxation around 20 K and inelastic neutron scattering
at room temperature was demonstrated seven years later
\cite{bu88}. A very recent Raman experiment \cite{wieder} showed
the temperature independence of $f_s(V)$ up to room temperature,
together with an excellent agreement of the shape of the function
from mechanical data with the one from Raman data.

\begin{figure}[b]
\begin{center}
\vspace*{-0.5cm}
\hspace*{0.0ex}\epsfig{file=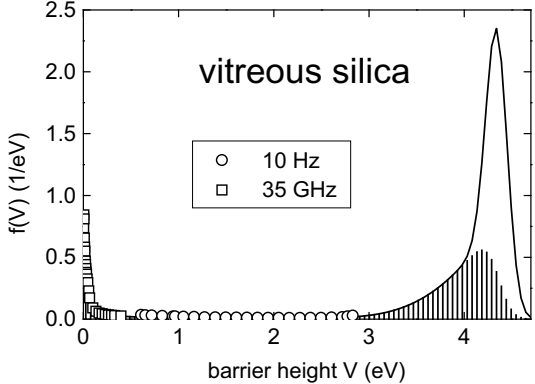,height=8.5cm,angle=-90}

\vspace*{0.0cm}
\end{center}
\caption{The barrier distribution function of vitreous silica,
together with torsion pendulum and Brillouin damping data.
References see text.}

\end{figure}

\begin{figure}[b]
\begin{center}
\vspace*{-0.5cm}
\hspace*{0.0ex}\epsfig{file=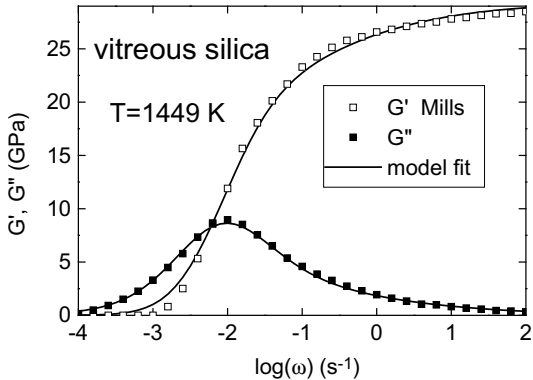,height=8.5cm,angle=-90}

\vspace*{0.0cm}
\end{center}
\caption{Dynamical mechanical data of vitreous silica at the glass
transition, together with the model fit (see text).}

\end{figure}

Of the many low-barrier mechanical data in the literature, we
include here merely one of the Brillouin damping experiments at 35
GHz \cite{vacher}. The range of higher barriers is covered by a
torsion pendulum measurement \cite{bruck} and a mHz experiment at
the glass transition itself \cite{mills}. Fig. 8 shows the barrier
distribution function fitting these data over the whole range.
Again, it turns out to be possible to fit the $\alpha$-relaxation
in terms of a gaussian for the primary barrier distribution
function $f_\alpha(V)$, ascribing the slow rise of the damping
towards the glass transition to secondary relaxation (the shaded
area in Fig. 8).

Fig. 9 shows the fit of the glass transition data \cite{mills},
calculated with $T_g=1460$ K, $G=31\ $GPa, $w_\alpha=0.55$ and
$\Delta_\alpha=0.263\ $eV. If one takes the value $T_g-T_0=850\ $K
and its rather large error bars from the shift factors determined
in \cite{mills}, one calculates $\alpha_s=6\pm 4$ from eq.
(\ref{fin}).

\begin{figure}[b]
\begin{center}
\vspace*{-0.5cm}
\hspace*{0.0ex}\epsfig{file=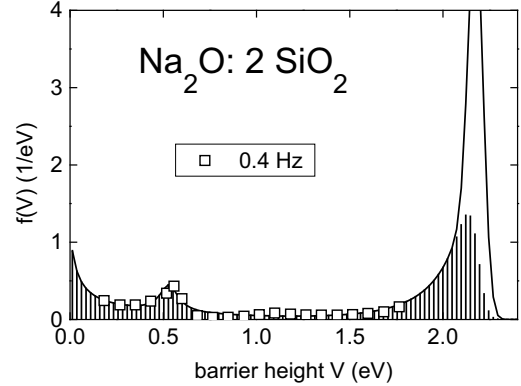,height=8.5cm,angle=-90}

\vspace*{0.0cm}
\end{center}
\caption{The barrier distribution function of Na$_2$O:2SiO$_2$,
together with torsion pendulum data. References see text.}

\end{figure}

\begin{figure}[b]
\begin{center}
\vspace*{-0.5cm}
\hspace*{0.0ex}\epsfig{file=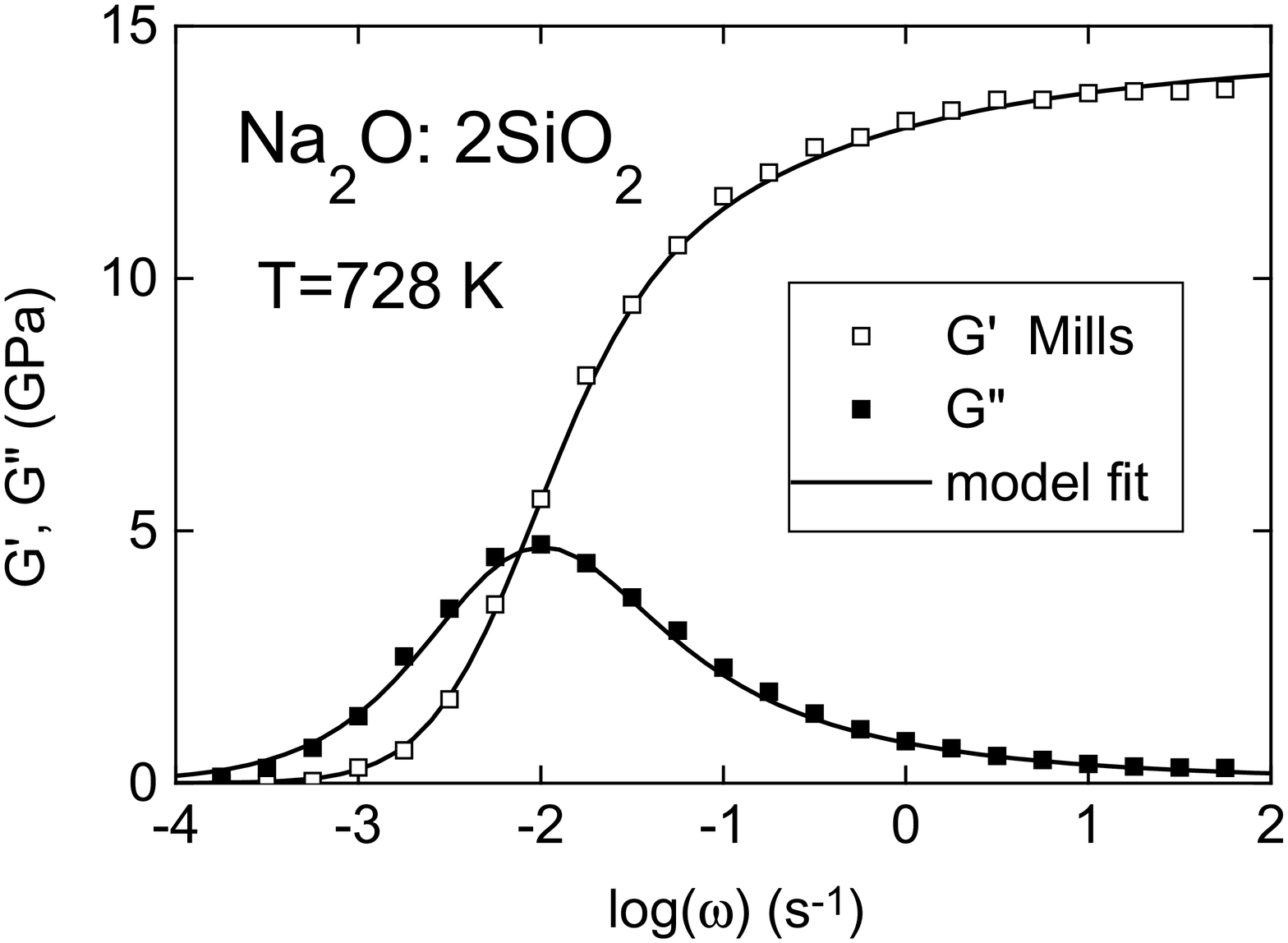,height=8.5cm,angle=-90}

\vspace*{0.0cm}
\end{center}
\caption{Dynamical mechanical data of Na$_2$O:2SiO$_2$ at the
glass transition, together with the model fit (see text).}

\end{figure}

It is interesting to compare vitreous silica to a sodium silicate
glass, Na$_2$O:2SiO$_2$, which has a much lower glass temperature
$T_g=733\ $K and is much more fragile \cite{mills}, with
$T_g-T_0=267\ $K. Fig. 10 shows the barrier distribution function,
together with torsion pendulum data \cite{graham} at 0.4 Hz. The
$\alpha$-peak region is again fitted to the mHz data \cite{mills},
shown in Fig. 11, with $G=19\ $GPa, $w_\alpha=0.39$ and
$\Delta_\alpha=0.092\ $eV. The shift factors \cite{mills} require
$\alpha_s=13\pm 1.5$.

\section{Results and discussion}

\subsection{Results for the flow process}

The preceding section showed for all three examples a good fit of
the $\alpha$-relaxation in terms of a narrow gaussian primary
barrier distribution function, together with a bigger or smaller
contribution of the secondary relaxation. The contribution of the
secondary relaxation to the $\alpha$-process was biggest in the
strong glass former silica and smallest in the very fragile case
of polystyrene. In all three cases, excellent fits of the time or
frequency dependence of the $\alpha$-process could be obtained
(see Figs. 7, 9 and 11). Simultaneously, the temperature
dependence of the shift factors for $\tau_\alpha$ could be
described accurately by the temperature slope $\alpha_s$ of the
rise of the secondary relaxation above $T_g$ in eq.
(\ref{fsrise}).

Table I compiles the fit parameters for the $\alpha$-process
determined in the preceding section.

\bigskip
\begin{center}
Table I: Model parameters for the $\alpha$-process
\end{center}
\begin{center}
\begin{tabular}{|l|c|c|c|}
\hline     substance       & polystyrene & SiO$_2$ &
Na$_2$O:2SiO$_2$
\\ \hline      $T_g$ (K)       &     372     & 1460 & 733       \\
    $G(T_g)$ (GPa)    &    1.69     &   31    &       19       \\
      $w_\alpha$      &    0.54     &  0.54   &      0.39      \\
 $\Delta_\alpha$ (eV) &    0.107    &  0.263  &     0.092      \\
      $\alpha_s$      & 6.9$\pm$0.1 & 6$\pm$4 &   13$\pm$1.5   \\
      \hline
\end{tabular}
\bigskip
\end{center}

In all three cases, the relative weight $w_\alpha$ of the
$\alpha$-process is not very far from 1/2. The $\alpha$-process
seems to occur when the secondary relaxations reduce the shear
modulus to about half its infinite frequency value, i.e. when the
secondary relaxational shear response equals the elastic one. The
situation reminds one of the definition of the Maxwell time
$\tau_M=\eta_0/G$, the time when the flow response equals the
elastic one. Thus one could think of a generalized Maxwell
criterion for the onset of the flow process, namely when the shear
response by relaxation is the same as the elastic shear response.
Intuitively, this is plausible: the time at which a macroscopic
shear stress for a given strain decays to half its initial value
should correspond to the lifetime of the microscopic shear pattern
of the glass former, because the microscopic stresses will be
expected to decay on the same time scale as the macroscopic ones.
But a decay of the microscopic stresses necessarily implies a
coupling between different relaxation centers in the glass. Thus
one would have to go beyond the simple idea of uncoupled
relaxation centers to understand the flow process \cite{ngai}.

In all three cases, the width $\Delta_\alpha$ is relatively
narrow, less than one tenth of the barrier $V_\alpha(T_g)$ itself,
in the third case even less than the twentieth part.

Finally, the coefficient $\alpha_s$ for the temperature rise of
the secondary barrier distribution function $f_s(V)$ above $T_g$,
related to the fragility of the glass former, is not so very
different for the three cases. There is a factor of two between
silica and the silicate glass, but taking the differences in the
glass temperature into account, one sees that the relative rise
per Kelvin is the same for the two glasses, admittedly within
large error bars. The difference is rather in $f_s(V)$ itself;
obviously, relaxing entities form much more easily and with much
lower barriers in the multiply broken network of the sodium
silicate glass than in the continuous random network of silica,
making the glass temperature a factor of two lower.

The large difference in fragility between silica and polystyrene
is not so much due to a difference in $\alpha_s$, but rather in
the product $Vf_s(V)$ at $T_g$ (see Figures 6 and 8), which is
about 0.3 in polystyrene and about 2.5 in silica. This product
enters into eq. (\ref{fin}) for the Vogel-Fulcher temperature
$T_0$. In this view, a substance is fragile when it has a low
density of secondary relaxing entities at the relaxation time of
the flow process. Naturally, the fast rise of the number of
secondary relaxing entities above $T_g$ remains a necessary
condition.

The coefficient $\alpha_s=6.9\pm 0.1$ for polystyrene in Table I,
obtained from the temperature dependence of the shift factors (see
Fig. 7), agrees within experimental error with the values
$\alpha_s=7.7\pm 1$ and $\alpha_s=10\pm 3$ determined from neutron
and Brillouin data above $T_g$ (see Fig. 5).

\subsection{Discussion}

Here is the proper place to remind the reader that with all the
presented equations and fits, the glass transition is not really
explained. The Gilroy-Phillips formalism supplies only a
reformulation of well-known equations in terms of a barrier
distribution function. It is a new way to look at the data, an
encouragement to seek explanations in a new direction,
complementary to attempts to understand the glass transition from
the liquid side \cite{gotz}, because it starts from a description
of the relaxation in the glass phase. But the present work does
neither explain the fast rise of the secondary relaxation above
$T_g$, nor does it supply a quantitative explanation why the shear
modulus breaks down completely when it is halved by the secondary
relaxation. It merely helps to quantify and to visualize these
experimental facts.

The concept of the barrier distribution function is based on the
idea of independent thermally activated relaxation processes in
disordered surroundings. As long as this idea applies without any
restriction, one should find a temperature-independent barrier
distribution function, as one does indeed in the glass phase of
polystyrene and silica. The quantitative comparison of
measurements at different temperatures and frequencies with the
equations of Section II.C enables a much more stringent check of
the Arrhenius concept than a mere Arrhenius temperature shift of a
broad relaxation peak with frequency \cite{arnold,hunk,heijboer}.
From the few such stringent checks reported so far, it is already
clear that the concept does not always work perfectly well in the
glass phase. In BPA-PC, another amorphous polymer, there is Raman
evidence \cite{surov} for an increase of the secondary barrier
distribution function $f_s(V)$ with temperature below $T_g$ , even
though the $\gamma$-relaxation peak shift is perfectly
Arrhenius-like \cite{floudas}. There seem to be more such cases
\cite{alex}. This is yet another question calling for a closer
investigation.

The outcome of the fits of the flow process in terms of the
primary barrier distribution function $f_\alpha(V)$ shows once
more the collectivity of the flow process. These are clearly not
independent thermally activated relaxation processes; otherwise
$V_\alpha$ would not shift with temperature. There has been an
attempt \cite{dyre} to explain the temperature dependence of
$V_\alpha(T)$ in terms of a proportionality to $G(T)$ ("flow by
shoving"). However, as pointed out in Section III.C, one finds a
Gr\"uneisen $\Gamma_g=3.8$ for $G(T)$ in polystyrene. This is a
factor of seven too weak to explain the temperature shift of
$V_\alpha(T)$.

In any case, the description of the flow process in terms of
$f_\alpha(V)$ in principle does not prejudice anything. One can
always return from $f_\alpha(V)$ to the conventional description
\cite{ferry} in terms of $H(\tau)$ via eq. (\ref{rheo}). What
might be questionable is the specific subdivision into primary and
secondary relaxation. Here, it was decided to fit the primary
barrier distribution function $f_\alpha(V)$ with a gaussian. On
the basis of the data, this choice can be only justified for
polystyrene; in the other two cases, one could choose differently.
The conclusion that the shear modulus is reduced to about half its
infinite frequency value by the secondary relaxation alone would
then no longer be valid for these two cases.

Independent of the choice of a specific function for the primary
barrier distribution function, the new description is much more
convenient than the old one if one wants to compare different
temperatures. It does not only allow for a stringent check of the
Arrhenius behaviour in the glass phase, but it also holds the
promise to provide a deeper understanding of the glass transition
riddles \cite{edi}. As shown in Section II.E, the stretching of
the flow process is quantitatively related to the width of the
$\alpha$-peak. More important, the Vogel-Fulcher behavior is an
inherent feature of the generic case of a constant secondary
barrier distribution function, provided it has a constant
temperature rise. But one can as well understand deviations
\cite{stickel1,stickel2} from the Vogel-Fulcher behavior in terms
of maxima and minima of the secondary barrier distribution
function. In particular, one must expect deviations from a single
Vogel-Fulcher law when $V_\alpha(T)$ sweeps through a secondary
relaxation peak, as one indeed observes \cite{stickel3}.

\section{Summary and conclusions}

The Gilroy-Phillips model \cite{gilroy} for secondary relaxation
in glasses, based on the Arrhenius-Kramers \cite{kramers} picture
of thermally activated jumps over energy barriers, has been
extended to describe the flow process (primary or
$\alpha$-process). The proper treatment of uncoupled thermally
activated relaxation events in disordered surroundings yields a
temperature-independent barrier distribution function $f(V)$. The
temperature independence holds as long as one has a constant
number of independent relaxation centers. The function has to
integrate to 1 in order to bring the shear modulus down to zero, a
very convenient sum rule. As it turns out, one can use the
formalism to describe any kind of relaxation, admitting a
temperature dependence of the barrier distribution function. Thus
one can separate the trivial Arrhenius temperature dependence from
nontrivial temperature changes of the relaxation.

The relaxation below the glass temperature $T_g$, in the frozen
energy landscape of the glass, should be describable in terms of a
temperature-independent barrier distribution function (in some
cases like polystyrene, CKN and silica \cite{surov,wieder} it
works, in others it does not \cite{surov,alex}), but the flow
process is most certainly of a different kind. Therefore one has
to postulate a secondary barrier distribution function $f_s(V)$
for the secondary relaxation, and an additional primary barrier
distribution function $f_\alpha(V)$ for the flow process. This
additional primary function shifts its maximum to lower values
with increasing temperature and serves simultaneously as a
relatively sharp cutoff for the secondary relaxation. This
property supplies a quantitative basis for the puzzling relation
between the fragility and the fast rise of the picosecond process
above $T_g$ \cite{angnew}. It explains the unusual temperature
dependence of the flow process in terms of the temperature rise of
the secondary relaxation, and  might even shed some light on the
detailed temperature dependence of the flow process in specific
substances \cite{stickel1,stickel2,stickel3}.

Reformulating the classical rheological equations \cite{ferry},
experimental literature data in polystyrene, vitreous silica and a
sodium silicate glass could be fitted in terms of the barrier
distribution functions. The results corroborate the earlier
conclusion \cite{surov} of a temperature-independent secondary
barrier distribution function in the glass phase of polystyrene,
but show a strong increase of the secondary relaxation with
increasing temperature above $T_g$. Taking the flow process as a
gaussian barrier distribution function, the fits show the onset of
the flow process more or less at the point where the secondary
relaxation reduces the shear modulus to half its infinite
frequency value. This was not only found in polystyrene, but in
the two other cases as well. It remains to be seen whether such a
generalized Maxwell criterion for the onset of the flow process is
a general property of glass formers. In polystyrene, the
temperature coefficient of the fast rise of $f_s(V)$ above $T_g$
calculated from the fragility coincided within experimental error
with the fast rise seen in neutron and Brillouin experiments.

The findings suggest a rapid change of the energy landscape above
$T_g$, the landscape getting more and more rugged with more and
more minima and saddle points as the temperature rises, thus also
explaining the entropy crisis at the Kauzmann temperature. This
does not really answer the central questions around the glass
transition, but focusses the attention on two points: (i) what is
the possible mechanism of formation of an increasing number of
relaxing entities with increasing temperature above $T_g$ (ii) why
does the flow process set in when the integrated relaxational
response to an external shear stress equals the elastic one.

{\bf Acknowledgements}: Helpful discussions with A. Wischnewski,
E. W. Fischer and K. L. Ngai are gratefully acknowledged.

\newpage

\begin{center}

{\bf Figure Captions}

\end{center}

\bigskip

\noindent{\bf Fig. 1:} Asymmetric double well potential with
barrier height $V$ and asymmetry $\Delta$ as a function of a
generalized coordinate.

\bigskip

\noindent{\bf Fig. 2:} Secondary barrier distribution function
$f_s(V)$ (the shaded area), together with the cutoff by the
$\alpha$-relaxation peak (schematic).

\bigskip

\noindent{\bf Fig. 3:} Secondary barrier distribution function
$f_s(V)$ calculated from literature data of the mechanical damping
of amorphous polystyrene in the glass phase at different
frequencies. References see text. The line is a fit; the same fit
is also shown in Figs. 4, 5 and 6.

\bigskip

\noindent{\bf Fig. 4:} Secondary barrier distribution function
$f_s(V)$ calculated from literature data of neutron and Raman
scattering from amorphous polystyrene in the glass phase.
References see text.

\bigskip

\noindent{\bf Fig. 5:} Secondary barrier distribution function
$f_s(V)$ calculated from literature data of neutron and Brillouin
data of amorphous polystyrene, both in the glass phase and above
the glass temperature $T_g$. References see text.

\bigskip

\noindent{\bf Fig. 6:} Secondary barrier distribution function
$f_s(V)$ calculated from literature mechanical damping and neutron
data of amorphous polystyrene up to the glass transition. The peak
at the end shows the gaussian $f_\alpha(V)$ describing the
$\alpha$-process at the glass temperature $T_g$. References see
text.

\bigskip

\noindent{\bf Fig. 7:} Creep data of the $\alpha$-process in
polystyrene at the glass transition (reference see text), together
with a fit in terms of $f(V)$.

\bigskip

\noindent{\bf Fig. 8:} The barrier distribution function of
vitreous silica, together with torsion pendulum and Brillouin
damping data. References see text.

\bigskip

\noindent{\bf Fig. 9:} Dynamical mechanical data of vitreous
silica at the glass transition, together with the model fit (see
text).

\bigskip

\noindent{\bf Fig. 10:} The barrier distribution function of
Na$_2$O:2SiO$_2$, together with torsion pendulum data. References
see text.

\bigskip

\noindent{\bf Fig. 11:} Dynamical mechanical data of
Na$_2$O:2SiO$_2$ at the glass transition, together with the model
fit (see text).

\bigskip

\end{multicols}

\end{document}